\documentclass[a4paper,notitlepage,12pt]{article}

\begin{document}
\title{Effects of Marangoni numbers on thermocapillary drop migration:
constant for quasi-steady state?}
\author{ Zuo-Bing Wu\footnotemark$^*$[1] and Wen-Rui Hu$^+$\\
State Key Laboratory of Nonlinear Mechanics$^*$, \\
 National Microgravity Laboratory$^+$,
 Institute of Mechanics,\\
  Chinese Academy of Sciences, Beijing 100190, China}
 \maketitle

\footnotetext[1]{Corresponding author. Tel:. +86-10-82543955;
fax.: +86-10-82543977. \\
Email addresses: wuzb@lnm.imech.ac.cn (Z.-B. Wu)}

\newpage
\begin{abstract}
The overall {\it steady}-state energy balance with two phases in a
flow domain requires that the change in energy of the domain is
equal to the difference between the total energy entering the
domain and that leaving the domain. From the condition, the
integral thermal flux across the surface is studied for a {\it
steady} thermocapillary drop migration in a flow field with
uniform temperature gradient at small and large Marangoni
(Reynolds) numbers. The drop is assumed to have only a slight
axisymmetric deformation from a sphere. It is identified that a
conservative/nonconservative integral thermal flux across the
surface in the {\it steady} thermocapillary drop migration at
small/large Marangoni (Reynolds) numbers. The conservative flux
confirms the assumption of {\it quasi-steady} state in the
thermocapillary drop migration at small Marangoni (Reynolds)
numbers.  The nonconservative flux may well result from the
invalid assumption of {\it quasi-steady} state, which indicates
that the thermocapillary drop migration at large Marangoni
(Reynolds) numbers cannot reach {\it steady} state and is thus a
{\it unsteady} process.


\end{abstract}

\newpage
\section{Introduction}
The motion of a drop or bubble in the microgravity environment
embedded in an immiscible mother liquid with a uniform temperature
gradient is termed as thermocapillary migration of the drop or
bubble, which is a very interesting topic for both fundamental
theory and engineering application[1].
 Young et al(1959) carried out an initial
study in this area, called as YGB model[2],
 and gave an analytical prediction on its
migration speed in the limit case of zero Reynolds(Re) and zero
Marangoni(Ma) numbers, and a series of theoretical analyses,
numerical simulations and experimental investigations on this
subject were carried out ever since. Subramanian(1981)[3] and
Crespo et al(1996)[4] extended the YGB results to small Ma numbers
and obtained analytical results in series expansion of Ma numbers.
With consideration of thermal boundary layer, the analytical
results for migration speed of a bubble at large Ma (Re)
numbers[5,6] agree well with the corresponding results of steady
state numerical simulations[7,8,4] and experimental studies[9].

Although the thermocapillary bubble migration processes are
understood very well, the behavior of thermocapillary drop
migration appears rather complicated due to the transfer of
momentum and energy though the interface of two-phase fluids. For
the migration of a drop, on the one hand, the experimental result
of the migration speed at small drops $11 \pm 1.5 \mu m$ in
diameter obtained by Braun et al(1993)[10] agrees with the YGB
model. On the other hand, another experiment for larger drops with
diameters ranging from 0.69 to 2.38 $mm$ performed by
Wozniak(1991)[11] shows that the migration velocities are smaller
than those given by the YGB linear prediction. Afterward more
attention has been paid to thermocapillary drop migration for
large Ma (Re) numbers. Hadland et al(1999)[9] carried out
experiments based on Fluorinert liquid FC-75 and 10cst silicone
oil as the drop phase and continuous phase during the NASA Space
Shuttle mission with the maximal Ma(Re) number up to 3300(49.1).
It was shown that the drop migration speed nondimensionalized by
the YGB velocity decreased as  the Ma number increased and the
global migration process exhibited an unsteady nature. To further
observe the variation trend of drop migration velocity with
increasing Ma number, Xie et al(2005)[12] adopted Fluorinert
liquid FC-75 and 5cst silicone oil as the drop and the continuous
phases, respectively, and performed experiments in the Chinese
spacecraft ShenZhou-4. The experimental investigation was
completed for several ranges of large Ma (Re) numbers, where the
drop didn't reach the steady state in the migration process, and
the maximal Ma(Re) number reached 5525(302.6). It was also
observed that the non-dimensional drop migration velocity
decreased as the Ma numbers increased. However, from the
theoretical analysis for the large Ma (Re) numbers, it was
reported [13] that the migration speed of a drop increased with
increasing Ma number, which is in qualitative agreement with the
corresponding numerical simulation[14]. Both the theoretical
analysis and numerical simulation are based on the assumptions of
quasi-steady state and non-deformation of the drop. The above
qualitative difference between experimental observations and
theoretical/numerical results may result from the quasi-steady
state or non-deformation assumptions of the drop in the model.
Moreover, Herrmann et al\cite{18} and Wu et al\cite{20} adopted
respectively the numerical methods to investigate the
thermocapillary motion of deformable and non-deformable drops and
indicated that the assumption of quasi-steady state was not valid
for large Ma numbers. Therefore, the thermocapillary drop
migration at large Ma (Re) numbers is still a topic to be further
studied with emphasis laid on its physical mechanism.

To address the discrepancies between experimental and
theroretical/numerical results, in this paper, our effects are
focused on the assumption of quasi-steady state in the process of
thermocapillary drop migration. The drop may have only a slight
axisymmetric deformation from a sphere. By using the asymptotic
expansion method, we investigate the continuity of integral
thermal flux across the surface based on the overall steady-state
energy balance in the flow domain, and analyze the existence of
quasi-steady migration of the drop at zero, small and large Ma
(Re) numbers.

\section{Models and quasi-steady state assumption}
Consider the thermocapillary migration of a spherical drop of
radius $R_0$, density $\gamma \rho$, dynamic viscosity $\alpha
\mu$, thermal conductivity $\beta k$, and thermal diffusivity
$\lambda \kappa$ in a continuous phase fluid of infinite extent
with density $\rho$, dynamic viscosity $\mu$, thermal conductivity
$k$, and thermal diffusivity $\kappa$ under a uniform temperature
gradient $G$. The change rate of the interfacial tension between
the drop and the continuous phase fluid with temperature is
denoted by $\sigma_T$. Axisymmetric energy equations for the
continuous phase and for the fluid in the drop in a laboratory
coordinate system denoted by a bar are written as follows
\begin{equation}
\begin{array}{l}
\frac{\partial{\bar{T}}}{\partial t} + \bar{\bf v} \bar{\nabla} \bar{T}= \kappa \bar{\Delta} \bar{T},\\
\frac{\partial{\bar{T'}}}{\partial t} + \bar{\bf v'} \bar{\nabla} \bar{T'}= \lambda \kappa \bar{\Delta} \bar{T'},
\end{array}
\end{equation}
where $\bar{\bf v}$ and $\bar{T}$ are velocity and temperature,
and a prime denotes quantities inside the drop. The solutions of
Eq. (1) have to satisfy the boundary conditions at infinity
\begin{equation}
{\bar T}_\infty \to  T_0 + G{\bar z},
\end{equation}
where $T_0$ is the undisturbed temperature of the continuous phase
and the boundary conditions at the interface $({\bar r}_b,{\bar
z}_b)$ of the two fluids
\begin{equation}
\begin{array}{l}
{\bar T}({\bar r}_b,{\bar z}_b,t) ={\bar T'}({\bar r}_b,{\bar z}_b,t),\\
\frac{\partial{{\bar T}}}{\partial n}({\bar r}_b,{\bar z}_b,t) =
\beta \frac{\partial{{\bar T'}}}{\partial n}({\bar r}_b,{\bar
z}_b,t),
\end{array}
\end{equation}
where ${\bf n}$ is a unit vector normal to the interface. In what
follows, the undisturbed temperature $T_0$ is reduced for
simplicity.

In general, the surface tension decreases with the increasing of
the local temperature. For a temperature field with its gradient
in the ${\bar z}$ direction, the generated surface tension force
is a net force along the surface and the droplet starts to move
towards the warm side under the action of net force. When the net
force acting on the drop at the flow direction is zero, the
thermocapillary drop migration reaches a stable process. However,
due to the variation of physical parameters with the ambient
temperature, the migration process may not reach stable state.
Only when
 the migration is sufficiently slow that the order of
 relevant time scale for the transport process to generate
 stable velocity and temperature fields is
 smaller than that for the drop to move an appreciable distance,
 the assumption of the quasi-steady state is valid.
It means that after experiencing an initial unstable migration
process, the drop migration may reach a steady state at the time
$t_0$ and the position ${\bf r}_0=z_0 {\bf k}$, i.e., migrating
with a constant drop migration speed $V_{\infty}$. Using the
coordinate transformation from the laboratory coordinate system to
a coordinate system moving with the drop velocity $V_{\infty}$
\begin{equation}
\begin{array}{lll}
\bar{\bf r} = {\bf r} + {\bf r}_0 + V_{\infty}(t-t_0) {\bf k},&
\bar{\bf v}(\bar{\bf r},t)
= {\bf v}({\bf r}) + V_{\infty} {\bf k},& \bar{T}(\bar{\bf r},t )= T({\bf r}) + G[z_0+V_{\infty}(t-t_0)],\\
&  \bar{\bf v'} (\bar{\bf r},t)  = {\bf v'}({\bf r}) + V_{\infty}
{\bf k},& \bar{T'}(\bar{\bf r},t ) = T'({\bf r})  + G[z_0 +
V_{\infty}(t-t_0)],
\end{array}
\end{equation}
 the problem (1) can be formulated as
\begin{equation}
\begin{array}{l}
G V_{\infty} + {\bf v} \nabla T= \kappa \Delta T,\\
G V_{\infty} + {\bf v'} \nabla  T'= \lambda \kappa \Delta T'.
\end{array}
\end{equation}
The details of the transformation are given in Appendix 1. By
taking the radius of the drop $R_0$, the velocity $v_0=-\sigma_T G
R_0/\mu$ and $GR_0$ as the reference quantities to make the
coordinates, velocity and temperature dimensionless, energy
equations (5) combined with the continuous equations can be
written in the following dimensionless form in a spherical
coordinate system ($r,\theta$)
\begin{eqnarray}
V_{\infty}+\nabla \cdot ({\bf v}T) = \frac{1}{Ma} \Delta T,\\
V_{\infty}+\nabla \cdot ({\bf v'}T') =\frac{\lambda}{Ma} \Delta
T', \label{1}
\end{eqnarray}
where ${\bf v}=(u,v)$ and Marangoni number is defined as
\begin{equation}
Ma=\frac{v_0R_0}{\kappa}.
\end{equation}
By using the transformation (4), the boundary conditions (2) and
(3) can be respectively written in the form of dimensionaless as
follows
\begin{equation}
T \to r \cos \theta, {\rm as} \  r \to \infty
\end{equation}
 at places far away from the drop and
\begin{eqnarray}
T(r_0,\theta) =T'(r_0,\theta),\\
\frac{\partial{T}}{\partial n}(r_0,\theta) = \beta
\frac{\partial{T'}}{\partial n}(r_0,\theta)
\end{eqnarray}
at the interface of the two fluids. Thus, once the drop migration
reaches a steady state, the above problem in the laboratory
coordinate system can be described by the steady energy equations
(6)(7) with the boundary conditions (9)(10)(11) in the coordinate
system moving with the drop velocity. This implies the overall
steady-state energy balance with two phases in the flow domain in
the co-moving frame of reference.

\section{Integral thermal flux across the drop surface under the quasi-steady state assumption}

In general, for a two-phase flow, as is the case in the present
problem, if the quasi-steady state assumption is valid, the
solutions of the problem not only satisfy the differential energy
equations with boundary conditions, but also the overall
steady-state energy with two phases in the flow domain under
integral boundary conditions keeps balance. However, if the
quasi-steady state assumption is invalid,  the overall
steady-state energy with two phases in the flow domain under
integral boundary conditions is not balanced. This means the
solutions of the problem also cannot satisfy the differential
energy equations with boundary conditions. Thus, to confirm
whether the thermocapillary drop migrations at different Ma (Re)
numbers are always in the quasi-steady state processes, we may
analyze the overall steady-state energy of two phases in the flow
domain in the co-moving frame of reference under integral boundary
conditions.

For the thermocapillary drop migration, to get the overall
steady-state energy transport of two phases in the flow domain in
the co-moving frame of reference, we have to integrate Eq. (6) and
Eq. (7) in the continuous phase domain $(r\in
[r_0,r_{\infty}],\theta\in[0,\pi])$ and within the drop region
$(r\in [0,r_0],\theta\in[0,\pi])$ as
\begin{eqnarray}
\int_{r_0}^\infty \int_0^\pi V_{\infty} dV+\int_{r_0}^\infty
\int_0^\pi \nabla \cdot ({\bf v}T) dV
= \frac{1}{Ma} \int_{r_0}^\infty \int_0^\pi \Delta T dV,\\
\int_0^{r_0} \int_0^\pi V_{\infty} dV+\int_0^{r_0} \int_0^\pi
\nabla \cdot ({\bf v'}T') dV=\frac{\lambda}{Ma} \int_0^{r_0}
\int_0^\pi \Delta T' dV, \label{1}
\end{eqnarray}
and Eq. (10) and Eq. (11) at the drop surface
$(r=r_0,\theta\in[0,\pi])$ as
\begin{eqnarray}
\int_0^\pi T(r_0,\theta) dS=\int_0^\pi T'(r_0,\theta) dS,\\
\int_0^\pi \frac{\partial{T}}{\partial n}(r_0,\theta) dS= \beta
\int_0^\pi \frac{\partial{T'}}{\partial n}(r_0,\theta) dS,
\end{eqnarray}
where $dV=r^2 \sin\theta dr d\theta$ and $dS=r^2 \sin\theta
d\theta$.
 And
then transforming the volume integration of Eq. (12) and Eq. (13)
in the flow domains to the surface integration over the droplet
surface and the surface at infinity in terms of the Gaussian
formula, we have
\begin{equation}
\frac{2 V_\infty}{3} (r^3_\infty-\frac{1}{2} \int_0^\pi r_0^3 \sin
\theta d \theta) + \oint (uT)|_{r_{\infty}} dS - \oint (uT)|_{r_0}
dS = \frac{1}{Ma} (\oint \frac{\partial T}{\partial
n}|_{r_{\infty}} dS - \oint
\frac{\partial T}{\partial n}|_{r_0} dS)\\
\end{equation}
and
\begin{equation}
\frac{V_\infty}{3} \int_0^\pi r_0^3 \sin \theta d \theta + \oint
({u'} {T'})|_{r_0} dS= \frac{\lambda}{Ma} \oint \frac{\partial
{T'}}{\partial n}|_{r_0} dS.
\end{equation}
 Using the zero normal velocity boundary condition at the
interface, we can derive
\begin{equation}
\begin{array}{ll}
\int_0^{\pi} \frac{\partial T}{\partial n}|_{r_0} r^2_0 \sin
\theta d\theta = & r_\infty^2 \int_0^{\pi} \frac{\partial
T}{\partial r}|_{r_\infty} \sin \theta d\theta - Ma r_\infty^2
\int_0^{\pi} (u T)|_{r_{\infty}}
 \sin \theta d\theta \\
 & - \frac{2 V_\infty Ma}{3}
(r^3_\infty-\frac{1}{2} \int_0^\pi r_0^3 \sin \theta d \theta)
\end{array}
\end{equation}
and
\begin{equation}
 \int_0^{\pi} \frac{\partial {T'}}{\partial
n}|_{r_0} r^2_0 \sin \theta d\theta= \frac{V_\infty Ma}{3 \lambda}
\int_0^\pi r_0^3 \sin \theta d \theta,
\end{equation}
where the outer normal vector at infinity is the radial coordinate
axis. Thus, Eq. (18) and Eq. (19) display integral thermal fluxes
across the drop surface obtained from the overall energy
transport. We assume that the drop has only a slight axisymmetric
deformation from a sphere
\begin{equation}
r_0=1+f(\theta),\ \ \ f \ll 1.
\end{equation}
For this case, Eq. (18) and Eq. (19) may be written as
\begin{equation}
\begin{array}{ll}
\int_0^{\pi} \frac{\partial T}{\partial n}|_{r_0} r^2_0 \sin
\theta d\theta = & r_\infty^2 \int_0^{\pi} \frac{\partial
T}{\partial r}|_{r_\infty} \sin \theta d\theta - Ma r_\infty^2
\int_0^{\pi} (u T)|_{r_{\infty}}
 \sin \theta d\theta \\
& - \frac{2 V_\infty Ma}{3} [r^3_\infty-1-\frac{3}{2} \int_0^\pi f
\sin \theta d \theta +O(f^2)]
\end{array}
\end{equation}
and
\begin{equation}
 \int_0^{\pi} \frac{\partial {T'}}{\partial
n}|_{r_0} r^2_0 \sin \theta d\theta= \frac{V_\infty Ma}{3 \lambda}
[2 + 3 \int_0^\pi f \sin \theta d \theta +O(f^2)].
\end{equation}
To next-to-leading (first) order in $f$, Eq. (21) and Eq. (22) may
finally be written as
\begin{equation}
\begin{array}{ll}
\int_0^{\pi} \frac{\partial T}{\partial n}|_{r_0} r^2_0 \sin
\theta d\theta = &r_\infty^2 \int_0^{\pi} \frac{\partial
T}{\partial r}|_{r_\infty} \sin \theta d\theta - Ma r_\infty^2
\int_0^{\pi} (u T)|_{r_{\infty}}
 \sin \theta d\theta \\
 &- \frac{2 V_\infty Ma}{3} (r^3_\infty-1-\frac{3}{2} \int_0^\pi f
\sin \theta d \theta)
\end{array}
\end{equation}
and
\begin{equation}
 \int_0^{\pi} \frac{\partial {T'}}{\partial
n}|_{r_0} r^2_0 \sin \theta d\theta= \frac{V_\infty Ma}{3 \lambda}
(2 + 3 \int_0^\pi f \sin \theta d \theta).
\end{equation}
 From
Eq. (14) and Eq. (15), we have integral boundary conditions across
the drop surface
\begin{eqnarray}
\int_0^\pi T|_{r_0} r^2_0 \sin \theta d\theta=\int_0^\pi T'|_{r_0} r^2_0 \sin \theta d\theta,\\
\int_0^\pi \frac{\partial{T}}{\partial n}|_{r_0} r^2_0 \sin \theta
d \theta= \beta \int_0^\pi \frac{\partial{T'}}{\partial n}|_{r_0}
r^2_0 \sin \theta d\theta.
\end{eqnarray}
Thus, for a quasi-steady state thermocapillary migration of the
drop with the slight deformation from the sphere, the overall
steady-state energy balance of two phases in the flow domain in
the co-moving frame of reference requires that the integral
thermal fluxes (23) and (24) at the drop surface obtained from the
overall energy transport are self-consistent with the integral
boundary condition (26). In the following, we will investigate the
self-consistency for the different Ma(Re) numbers.

\subsection{Conservative integral thermal flux across the drop
surface at zero Ma (Re) numbers}

In the case of zero Re ($Re=\frac{v_0 R_0}{\nu}$) and zero Ma
numbers, i.e. the YGB model, scaled velocity and temperature
fields
 of the continuous phase and within the drop in Eq. (6)
and Eq. (7) may be described\cite{2,19} as
\begin{equation}
\begin{array}{ll}
u&=-V_{\infty} \cos \theta (1-\frac{1}{r^3}),\\
v&=V_{\infty} \sin \theta (1+\frac{1}{2r^3}),\\
T&=(r  + \frac{1-\beta}{2+\beta} \frac{1}{r^2}) \cos \theta,\\
\end{array}
\end{equation}
and
\begin{equation}
\begin{array}{ll}
u'&=\frac{3}{2} V_{\infty} \cos \theta (1-r^2),\\
v'&=-\frac{3}{2} V_{\infty} \sin \theta (1-2r^2),\\
T'&= \frac{3}{2+\beta}r \cos \theta.
\end{array}
\end{equation}
Since $Ma=0$, using the temperature field in (27), we can derive
the following equality from Eq. (23) and Eq. (24)
\begin{equation}
\int_0^{\pi} \frac{\partial{T}}{\partial{n}}|_{r_0} r^2_0 \sin
\theta d \theta =\beta \int_0^{\pi} \frac{\partial {T'}}{\partial
n}|_{r_0} r^2_0 \sin \theta d\theta=0.
\end{equation}
Thus, under the quasi-steady state assumption, the integral
thermal flux across the drop surface at zero Re and zero Ma
numbers is conservative, which corresponds to the integral thermal
flux boundary condition (26). This implies the overall
steady-state energy balance of two phases in the flow domain in
the co-moving frame of reference. The quasi-steady state
assumption is valid.

\subsection{Conservative integral thermal flux across the drop
surface at small Ma (Re) numbers}

For small Re number, the velocity fields in Eq. (6) and Eq. (7)
may be described by the creeping flow. The general solutions of
the scaled flow field in the continuous phase and within the drop
are given by \cite{15,16} as
\begin{equation}
\begin{array}{ll}
u&=-V_{\infty} \cos \theta (1-\frac{1}{r^3})  -(1-\frac{1}{r^2})
\sum_{n=3}^{\infty} D_n r^{-n+1} P_{n-1}(\cos
\theta),\\
v&=V_{\infty} \sin \theta (1+\frac{1}{2r^3}) +\sum_{n=3}^{\infty}
D_n (\frac{-n+3}{r^{n-1}}-\frac{-n+1}{r^{n+1}}) C_n^{-1/2}(\cos
\theta) /\sin \theta
\end{array}
\end{equation}
and
\begin{equation}
\begin{array}{ll}
u'&=\frac{3}{2} V_{\infty} \cos \theta (1-r^2) +(\frac{1}{r^2}-1)
\sum_{n=3}^{\infty} D_n r^n P_{n-1}(\cos \theta),\\
v'&=-\frac{3}{2} V_{\infty} \sin \theta (1-2r^2)
-\sum_{n=3}^{\infty} D_n [nr^{n-2}-(n+2)r^n] C_n^{-1/2}(\cos
\theta)/\sin \theta,
\end{array}
\end{equation}
where $C_n^{-1/2}(\cos \theta)$ is the Gegenbauer polynomial of
order $n$ and degree $-\frac{1}{2}$, $P_{n-1}(\cos \theta)$ is the
Legendre polynomial of order $n$. $D_n$ is given as
\begin{equation}
D_n=-\frac{n(n-1)}{4(1+\alpha)} \int_0^{\pi}  C_n^{-1/2}(\cos
\theta) \frac{\partial T}{\partial \theta} d \theta.
\end{equation}
 And scaled temperature fields
 in the continuous phase and within the drop at the small Ma numbers are given in \cite{16} as
\begin{equation}
\begin{array}{ll}
T= &(r + \frac{1-\beta}{2+\beta} \frac{1}{r^2}) \cos \theta \\
& + \frac{1}{3 \lambda(2+\beta)^2(2+3 \alpha)} [\frac{\delta_1}{r}
+\frac{\delta_2}{r^4} +P_2(\cos \theta) (\frac{\delta_3}{r}
+\frac{\delta_4}{r^3} + \frac{2\delta_2}{r^4})] \epsilon
+O(\epsilon^2),\\
T'=& \frac{3}{2+\beta}r \cos \theta \\
& + \frac{1}{\lambda (2+\beta)^2(2+3\alpha)} [\delta'_1 +
\delta'_2 r^2 -\frac{3}{4} r^4 +P_2(\cos \theta) (\delta'_4 r^2
+\frac{3}{7} r^4)] \epsilon +O(\epsilon^2)
\end{array}
\end{equation}
where $\delta_1 = 2 [\lambda(1-\beta) -\beta(2+\beta)]$, $\delta_2
= -\frac{\lambda}{2}(1-\beta)$, $\delta_3 = -\lambda(4-\beta)$,
$\delta_4 = \frac{1}{7(3+2\beta)} [7\lambda
(8+5\beta-4\beta^2)-18\beta]$, $\delta'_1 = \frac{1}{12}
[6\lambda(1-\beta)-(8\beta^2 +20\beta +17)]$, $\delta'_2 =
\frac{1}{6} (2\beta+13)$, $\delta'_4 = -\frac{1}{21(3+2\beta)} [
7\lambda (7-\beta) + 9(3+4\beta)]$ and the small parameter is
$\epsilon=Ma$. Then, using the temperature field in (33), we
simplify Eq. (23) and Eq. (24) to
\begin{equation}
\begin{array}{lll}
\int_0^{\pi} \frac{\partial{T}}{\partial{n}}|_{r_0} r^2_0 \sin
\theta d \theta & = & -\frac{2}{3 \lambda(2+\beta)^2(2+3 \alpha)}
(\delta_1 +\frac{4\delta_2}{r^3_\infty}) \epsilon +
\frac{2(1-\beta)}{3(2+\beta)}(1-\frac{1}{r_\infty^3}) V_\infty
\epsilon \\
& & + V_\infty \int_0^\pi f \sin \theta d \theta  \epsilon + O(\epsilon^2)\\
& \approx & -\frac{2\delta_1}{3 \lambda (2+\beta)^2(2+3\alpha)}
\epsilon
+ \frac{2(1-\beta)}{3(2+\beta)} V_\infty \epsilon + V_\infty  \int_0^\pi f \sin \theta d \theta \epsilon + O(\epsilon^2)\\
& = & [\frac{2\beta}{3\lambda} V^0_\infty +
\frac{2(1-\beta)}{3(2+\beta)}
(V_\infty - V^0_\infty) + V_\infty \int_0^\pi f \sin \theta d \theta] \epsilon + O(\epsilon^2). \\
\end{array}
\end{equation}
and
\begin{equation}
\int_0^{\pi} \frac{\partial{T'}}{\partial{n}}|_{r_0} r^2_0 \sin
\theta d \theta =\frac{V_\infty}{3 \lambda} (2 +
3 \int_0^\pi f \sin \theta d \theta) \epsilon, \\
\end{equation}
where $V^0_\infty=\frac{2}{(2+\beta)(2+3 \alpha)}$ is the
migration velocity of the droplet at zero Re and zero Ma numbers.
From Eq. (34) and Eq. (35), we have
\begin{equation}
\begin{array}{ll}
\int_0^{\pi} (\frac{\partial{T}}{\partial{n}}|_{r_0} - \beta
\frac{\partial{T'}}{\partial{n}}|_{r_0}) r^2_0 \sin \theta d
\theta = & \frac{2}{3} (\frac{\beta}{\lambda} -
\frac{1-\beta}{2+\beta}) (V^0_\infty - V_\infty) \epsilon \\
&+ (1-\frac{\beta}{\lambda}) V_\infty \int_0^\pi f \sin \theta d
\theta \epsilon + O(\epsilon^2).
\end{array}
\end{equation}
 Since $\epsilon \to 0(V_\infty \to V^0_\infty)$ and $f \ll 1$, we have
\begin{equation}
\int_0^{\pi} \frac{\partial{T}}{\partial{n}}|_{r_0} r^2_0 \sin
\theta d \theta \approx \beta \int_0^{\pi}
\frac{\partial{T'}}{\partial{n}}|_{r_0} r^2_0 \sin \theta d
\theta.
\end{equation}
Thus, under the quasi-steady state assumption, the integral
thermal flux across the drop surface at small Re and small Ma
numbers is conservative, which corresponds to the thermal flux
boundary condition (26). This implies the overall steady-state
energy balance of two phases in the flow domain in the co-moving
frame of reference. The quasi-steady state assumption is valid.

\subsection{Nonconservative integral thermal flux across the drop
surface at large Ma (Re) numbers}
 For large Re number, the velocity fields in Eq. (6) and Eq. (7) can be
 described by potential flows and boundary layer flows\cite{17}.
 The scaled inviscid velocity field
 in the continuous phase and Hill's spherical vortex within the drop
 can be respectively written as
\begin{equation}
\begin{array}{l}
u=-V_{\infty}\cos \theta (1-\frac{1}{r^3}),\\
v=V_{\infty}\sin \theta (1+\frac{1}{2r^3})
\end{array}
\end{equation}
and
\begin{equation}
\begin{array}{l}
u'=\frac{3V_{\infty}}{2} \cos \theta (1-r^2),\\
v'=- \frac{3V_{\infty}}{2} \sin \theta (1-2r^2).
\end{array}
\end{equation}
Since Eq. (9) only gives the primary approximation of the
temperature field at infinity, we have to obtain an asymptotic
expansion of $T$ for the integration of Eq. (23). To determine the
asymptotic behavior of $T$ at $r \gg 1$, the analytical result of
outer temperature field in the continuous phase at the small
parameter $\epsilon=1/\sqrt{V_\infty Ma}$ is given in \cite{6}
  as
\begin{equation}
T=r \cos \theta + \int^{r}_\infty (v \sin \theta -u \cos \theta
-1)/u|_\Psi d\tilde{r} + o(1),
\end{equation}
where $\Psi[=\frac{1}{2}\sin^2 \theta(r^2-1/r)]$ is the
streamfunction of the continuous phase and the symbol "+" before
the integral is determined to preserve the monotonously increasing
trend of $T(r,0)$ with $r(>1)$ in the continuous phase. Using Eq.
(38), it can be derived as:
\begin{equation}
T=r \cos \theta + \int^{r}_\infty \frac{1}{\tilde{r}^3-1}
\frac{1-3 \sin^2 \theta}{2 \cos \theta}|_\Psi d\tilde{r} + o(1).
\end{equation}
 Replacing $\theta$ by $\Psi$ in Eq. (41), we
have
\begin{equation}
T=r \cos \theta + \int_r^\infty \frac{1}{\tilde{r}^3-1} \frac{3
\Psi / (\tilde{r}^2-1/\tilde{r})-1}{\pm
\sqrt{1-2\Psi/(\tilde{r}^2-1/\tilde{r})}}|_\Psi d\tilde{r} + o(1),
\end{equation}
where the symbol "$\pm$" in the integral depends on the value of
$\theta$ (the symbol $"+"/"-"$ corresponds to $\theta \in
[0,\pi/2)/[\pi/2,\pi)$).
 At $r \gg 1$, the
result (42) can be expressed as
\begin{equation}
\begin{array}{ll}
T &\approx r \cos \theta + \int_r^\infty \frac{1}{\tilde{r}^3}
\frac{3 \Psi / \tilde{r}^2-1} {\pm
\sqrt{1-2\Psi/\tilde{r}^2}}|_\Psi
d\tilde{r} + o(1)\\
 &= r \cos \theta -\frac{1}{2r^2} \cos \theta + o(1),
 \end{array}
\end{equation}
where $\Psi \approx \frac{1}{2}\sin^2 \theta r^2$.

Using the temperature field at the infinity in (43), we can
simplify Eq. (23) and Eq. (24) and derive
\begin{equation}
\begin{array}{ll}
\int_0^{\pi} \frac{\partial{T}}{\partial{n}}|_{r_0} r^2_0 \sin
\theta d \theta &= -\frac{1}{3\epsilon^2}
(1-\frac{1}{r^3_{\infty}}) +  \frac{1}{\epsilon^2} \int_0^{\pi} f
\sin \theta d \theta
 + o(\frac{1}{\epsilon^2}) \\
& \approx
-\frac{1}{3\epsilon^2}(1-3 \int_0^{\pi} f \sin \theta d \theta)
 \end{array}
\end{equation}
and
\begin{equation}
\int_0^{\pi} \frac{\partial{T'}}{\partial{n}}|_{r_0} r^2_0 \sin
\theta d \theta = \frac{1}{3 \lambda\epsilon^2}(2+3 \int_0^{\pi} f
\sin \theta d \theta).
\end{equation}
From Eq. (44) and Eq. (45), we have
\begin{equation}
\begin{array}{ll}
\int_0^{\pi} (\beta \frac{\partial{T'}}{\partial{n}}|_{r_0} -
\frac{\partial{T}}{\partial{n}}|_{r_0}) r^2_0 \sin \theta d \theta
 &=\frac{1}{3\epsilon^2} (1+ \frac{2 \beta}{\lambda})
 + (\frac{\beta}{\lambda}-1) \frac{1}{\epsilon^2} \int_0^{\pi} f \sin \theta d
 \theta\\
 &= \frac{1}{3} (1+ \frac{2 \beta}{\lambda}) V_{\infty}Ma +
 (\frac{\beta}{\lambda}-1) V_{\infty}Ma \int_0^{\pi} f \sin \theta d\theta.
 \end{array}
\end{equation}
Since both $\beta$ and $\lambda$ are positive and $f \ll 1$, we
have
\begin{equation}
\beta \int_0^{\pi} \frac{\partial{T'}}{\partial{n}}|_{r_0} r^2_0
\sin \theta d \theta \gg \int_0^{\pi}
\frac{\partial{T}}{\partial{n}}|_{r_0} r^2_0 \sin \theta d \theta.
\end{equation}
So, if the overall steady-state energy with two phases in the flow
domain under integral boundary conditions is balanced, Eq. (47)
should be reduced to Eq. (26), which seems impossible. It is
termed as a nonconservative integral thermal flux across the
surface for the steady thermocapillary drop migration at large Ma
(Re) numbers. This implies the overall steady-state energy
unbalance of two phases in the flow domain in the co-moving frame
of reference and indicates that the thermocapillary drop migration
at large Ma (Re) numbers cannot reach steady state. Thus, it is
clear that the invalid assumption of quasi-steady state for the
thermocapillary drop migration process is a reasonable explanation
for the nonconservative integral thermal flux across the drop
surface.

To analyze the thermal flux near the boundary, we write the
integrals of Eq. (44) and Eq. (45) in the discretization scheme as
follows
\begin{equation}
\int_0^{\pi} \frac{\partial{T}}{\partial{n}}|_{r_0} r^2_0 \sin
\theta d \theta = \sum_{i=1}^N \frac{\partial{T}}{\partial{n}}
(r_0,\theta_i) r^2_0 \sin \theta_i \Delta \theta <0\\
\end{equation}
and
\begin{equation}
\int_0^{\pi} \frac{\partial{T'}}{\partial{n}}|_{r_0} r^2_0 \sin
\theta d \theta = \sum_{i=1}^N \frac{\partial{T'}}{\partial{n}}
(r_0,\theta_i) r^2_0 \sin \theta_i \Delta \theta >0,
\end{equation}
where $\theta_i \in [0,\pi]$ and $\Delta \theta =\pi/N$. Since
$r^2_0 \sin \theta_i \geq 0$, we reach a conclusion that there
must be some interface points $\theta_i \in [0,\pi]$ where the
following equation holds
\begin{equation}
\frac{\partial{T}}{\partial{n}}(r_0,\theta_i)  < 0  <
 \frac{\partial{T'}}{\partial{n}}(r_0,\theta_i).
\end{equation}
or some interface points $\theta_i$ and $\theta_j \in [0,\pi]$
where the following equations hold
\begin{equation}
\begin{array}{l}
0< \frac{\partial{T}}{\partial{n}}(r_0,\theta_i) <
 \frac{\partial{T'}}{\partial{n}}(r_0,\theta_i),\\
\frac{\partial{T}}{\partial{n}}(r_0,\theta_j)   <
 \frac{\partial{T'}}{\partial{n}}(r_0,\theta_j)<0.
 \end{array}
\end{equation}
Physically, this means that near these points $\theta_i$ the
thermal energy is transferred from the interface to outside (the
surrounding fluid) as well as from the interface to inside (the
droplet) or near these points $\theta_i$/$\theta_j$ the
transference of thermal energy from outside/the interface to the
interface/outside is weaker/stronger than that from the
interface/inside to inside/the interface. On the one hand, if Eq.
(50) can satisfy the integral thermal flux boundary condition in
Eq. (26), thermal sources inside the interface will be introduced
to balance the transference of thermal energy. On the other hand,
if Eq. (51) can satisfy the integral thermal flux boundary
condition in Eq. (26), thermal sinks inside the interface or
thermal sources in the droplet will be introduced to decrease the
transference of thermal energy from the interface to outside or
increase the transference of thermal energy from inside to the
interface. Since there is absolutely no thermal sources or sinks
inside the interface or thermal sources in the droplet, the above
transport processes of thermal energy near the interface seem
impossible. It means that the thermal flux across the drop surface
is nonconservative.

\section{Conclusions}
In summary, from the condition of overall steady-state energy
balance with two phases in a flow domain, we have identified  a
conservative/nonconservative integral thermal flux across the
surface for a steady thermocapillary migration of a drop with a
slight axisymmetric deformation from a sphere in a uniform
temperature gradient at small/large Ma (Re) numbers. The
conservative integral thermal flux confirms the assumption of
quasi-steady state in thermocapillary drop migration at small Ma
(Re) numbers. The nonconservative integral thermal flux may well
result from the invalid assumption of quasi-steady state, which
indicates that the thermocapillary drop migration at large Ma (Re)
numbers cannot reach steady state and is thus a unsteady process.

\textbf{Acknowledgments} One of the authors (Z.B.W.) thanks the
National Science Foundation for
general support through the Grants No. 11172310 and
IMECH/SCCAS SHENTENG 1800/7000 research computing facilities for
assisting in the computation.

\newpage
\textbf{Appendix}

Based on the transformation between two cylindrical coordinate
systems ($\bar{r},\bar{z}$) and ($r,z$) of Eq. (4), we have
\begin{equation}
\begin{array}{l}
\bar{\nabla}|_t =\frac{\partial}{\partial \bar{r}}|_t {\bf i} + \frac{\partial}{\partial \bar{z}}|_t {\bf k}
=\frac{\partial}{\partial r}|_t {\bf i} + \frac{\partial}{\partial z}|_t {\bf k} = \nabla|_t,\\
\bar{\Delta}|_t =\frac{1}{\bar{r}} [\frac{\partial}{\partial \bar{r}} ({\bar{r}}\frac{\partial}{\partial \bar{r}})
+ \frac{\partial}{\partial \bar{z}} ({\bar{r}}\frac{\partial}{\partial \bar{z}}) ]|_t
=\frac{1}{r} [\frac{\partial}{\partial r} (r \frac{\partial}{\partial r})
+ \frac{\partial}{\partial z} ( r\frac{\partial}{\partial z}) ]|_t
= \Delta|_t.
\end{array}
\end{equation}
And for energy equation Eq. (1) of the continuous phase fluid, we
can also write its unsteady, convection and conductivity terms as
follows
\begin{equation}
\begin{array}{ll}
\frac{\partial \bar{T}}{\partial t}|_{\bar{\bf r}} &= \frac{\partial T}{\partial t}|_{\bar{\bf r}} + G V_{\infty}
=\frac{\partial T}{\partial r}|_t \frac{\partial r}{\partial t}|_{\bar{\bf r}}
+\frac{\partial T}{\partial z}|_t \frac{\partial z}{\partial t}|_{\bar{\bf r}}
+\frac{\partial T}{\partial t}|_{\bf r} \frac{\partial t}{\partial t}|_{\bar{\bf r}} + G V_{\infty}\\
&=\frac{\partial T}{\partial z}|_t (-V_{\infty}) +\frac{\partial T}{\partial t}|_{\bf r} + G V_{\infty}
=-  V_{\infty} \frac{\partial T}{\partial z} + G V_{\infty},\\
{\bar{\bf v}} \bar{\nabla} \bar{T}|_t & = ({\bf v} + V_{\infty} {\bf k}) \bar{\nabla} (T +GV_{\infty} t)|_t
=({\bf v} + V_{\infty} {\bf k}) \bar{\nabla} T|_t\\
&={\bf v} \nabla T  + V_{\infty} \frac{\partial T}{\partial z},\\
\bar{\Delta} \bar{T}|_t &= \bar{\Delta} (T +GV_{\infty} t)|_t = \bar{\Delta} T|_t =\Delta T,
\end{array}
\end{equation}
where $\frac{\partial r}{\partial t}|_{\bar{\bf r}}=\frac{\partial
r}{\partial t}|_{\bar{r}}=0$, $\frac{\partial z}{\partial
t}|_{\bar{\bf r}} =\frac{\partial z}{\partial
t}|_{\bar{z}}=-V_{\infty}$ and $\frac{\partial T}{\partial
t}|_{\bf r}=0$. Then, substituting Eq. (53) into Eq. (1) in the
laboratory coordinate system, we obtain Eq. (5) in a coordinate
system with its origin fixed at the drop center
\begin{equation}
\begin{array}{l}
G V_{\infty} + {\bf v} \nabla T= \kappa \Delta T.\\
\end{array}
\end{equation}
Similarly, we can also transform the energy equation within the
drop as above.

\end{document}